\documentclass[12pt,preprint]{aastex}
\begin{document}

\title{Resolved Mid-Infrared Emission Around AB Aur and V892 Tau with
Adaptive Optics Nulling Interferometric Observations$^{1}$}
\footnotetext[1]{The results presented here made use of the of MMT Observatory,
a jointly operated facility of the University of Arizona and the Smithsonian
Institution.}
\author{Wilson M. Liu$^{2}$,
Philip M. Hinz$^{2}$,
William F. Hoffmann$^{2}$,
Guido Brusa$^{2,3}$,
Doug Miller$^{2}$,
and Matthew A. Kenworthy$^{2}$}
\altaffiltext{2}{Steward Observatory, University of
Arizona, 933 N. Cherry Ave., Tucson, AZ, USA 85721, email: wliu@as.arizona.edu}
\altaffiltext{3}{Osservatorio Astrofisico di Arcetri, Largo Enrico Fermi 5,
50125 Florence, Italy}

\begin{abstract}
We present the results of adaptive optics nulling interferometric observations
of two Herbig Ae stars, AB Aur and V892 Tau.  Our observations at 10.3 $\mu$m
show resolved circumstellar emission from both sources.  Further analysis
of the AB Aur emission suggests that there is an inclined disk surrounding
the star.  The diameter of the disk is derived to be 24 to 30 AU with
an inclination of 45$\degr$ to 65$\degr$ from face-on, and a
major-axis PA of $30\degr \pm 15\degr$ (E of N).  Differences in the
physical characteristics between the mid-IR emission
and emission at other wavelengths (near-IR and millimeter), found in
previous studies, suggest a complex structure for AB Aur's circumstellar
environment, which may not be explained by a disk alone.  The similarity
in the observed size of AB Aur's resolved emission and that of another
Herbig Ae star, HD 100546, is likely coincidental, as their respective
evolutionary states and spectral energy distributions suggest significantly
different circumstellar environments.
\end{abstract}

\section{Introduction}
Over the past several years, numerous circumstellar disks have been
observed surrounding pre-main sequence (PMS) stars, including Herbig Ae (HAE)
stars.  HAE stars, the evolutionary precursors to intermediate-mass
main-sequence stars like Vega, show infrared (IR) emission in
excess of what is expected from their photosphere.  This emission was
thought to originate from optically thick, geometrically thin circumstellar
disks \citep{hill92,la92}, a model later modified to include disk flaring
in order to explain features in the observed spectral energy distributions
of the disks \citep{kh87,cg97}.  Alternative models also include dusty
envelopes of material or a combination of a disk and an envelope
\citep{hart93,miro99}.  Given the variety of models used to explain
the IR excess around PMS stars, observations of their circumstellar
environments are necessary to determine which of these models is most
representative of their true environment.

The HAE star AB Aurigae (A0; $d=144$ pc \citep{vdA98}) has been the subject
of many studies based upon observations at several different wavelengths.
The star has an estimated age of 2-5 Myr \citep{ms97,vdA98}. Near-IR (NIR)
emission from the AB Aur circumstellar region has been observed, probing
thermal emission in the inner AU of the disk using long-baseline
interferometry \citep{eis03,m-g01}. Both studies suggest the
presence of a slightly inclined distribution of dust with an empty or
optically thin inner region (i.e., a ring-like structure).
Another recent study in the NIR has detected scattered light from the disk
at greater separations (out to 580 AU) and finds the disk to have a small
inclination  \citep{f04}.
Observations in the mid-IR (MIR) suggest evidence for resolved circumstellar
material at 12 and 18$\mu$m at several tens of AU from the star
\citep{cj03,marsh}.  Longer wavelength
observations of AB Aur in the millimeter were shown to have spatially
resolved molecular line emission at a few hundred AU \citep{ms97}.
Reflection nebulosity has also been detected in the optical by
\citet{grady} which finds material out to 1300 AU and a disk inclination
of less than $45\degr$.

V892 Tau is a HAE star located in the Taurus-Auriga star forming
region, at a distance of about 140 pc \citep{elias}.  NIR speckle
interferometry of the star revealed an elongated structure with a PA of
$90\degr$. The source of emission is speculated to be either a highly inclined
disk or a bipolar outflow \citep{haas97}.

In this Letter, we present results of nulling interferometric observations
in the MIR of these two HAE stars, AB Aur and V892 Tau, in which
we have clearly resolved emission from both sources.  From
these observations, we infer and discuss the physical properties of the
circumstellar material in the AB Aur system.

\section{Observations and Data Reduction}
Observations were made in 2002 November and 2004 February at the 6.5 m MMT
Telescope on Mt. Hopkins, Arizona. The BracewelL Infrared Nulling
Cryostat (BLINC; \citet{hinz_phd}) provided supression of starlight, and the
Mid-Infrared Array Camera (MIRAC; \citet{hoff}) provided the final stop
for the two beams of the interferometer.  The 2002 run consisted of
preliminary observations of both science targets at 10.3$\mu$m
(10\% bandpass).  Four sets of 500 frames (each frame
with 0.5 s integration) were taken for each target, resulting in 2000 frames
for each source.  Images of a point-soure calibrator, $\alpha$ Ari, were
taken before and after the science observations.  Frames were sky subtracted
using off-source fields taken in between each set of frames.  Photometry was
extracted from each of the science frames, and each set of frames was
examined for the best instrumental null (Instrumental Null = Nulled Flux/
Unnulled Flux).  Source nulls were derived for each set of frames
(Source null = Instr. Null - Calibrator Null; see \citet{liu04} for a
full description of nulling calibration and diagnostics, as well as a
typical image from the BLINC-MIRAC camera).

Followup observations were made in 2004 at the MMT, also at 10.3 $\mu$m,
utilizing BLINC-MIRAC with the telescope's adaptive optics (AO) secondary
mirror. Since the deformable mirror is the telescope's
secondary mirror, there is no need for an intermediate set of reimaging and
correcting optics between the secondary mirror and science camera.
This has the benefit of optimizing throughput and
decreasing background emissivity in the MIR by avoiding the use of extra
warm optics.  The adaptive secondary also provides two major benefits
particular to our nulling observations: 1) The wavefront is stabilized,
allowing us to precisely tune the interferometer for the
best possible suppression of starlight;  2)  Fewer data frames need to be
taken since the wavefront stabilization allows us to precisely tune the
destructive interference.  These observations included 15 sets of 10 frames
(150 frames total) for AB Aur in destructive interference, and 3 sets of 10
frames for V892 Tau in destructive interference.  Frames of the objects
in constructive interference, as well as off-source sky frames were taken
in between each set of nulled images.

In order to probe the spatial distribution and orientation of the
circumstellar emission, the followup observations of AB Aur were taken at
several different rotations of the interferometer baseline relative to
the sky (see \citet{liu03} for a full description of this technique).
This was also attempted V892 Tau, however, poor observing conditions
on the night this object was observed allowed only limited followup.

\section{Results and Discussion}
The preliminary observations in 2002 show evidence for resolved emission
in both sources.  Source nulls derived from these non-AO observations
show the resolved flux surrounding each object to be at a level of 10 to
20\% of its full 10.3 $\mu$m flux (3-6 Jy for both sources).  Followup
observations in 2004 are described below.

\subsection{V892 Tau}
Due to the limited observations of V892 Tau in 2004, we make no
conclusions about the spatial distribution of emission around this object.
However, from our observations we are able to confirm the presence of
extended emission in the MIR first uncovered by our preliminary observations.
The extended emission is verified to be at a level of $11\% \pm 6\%$ (about 3 Jy;
$2 \sigma$ error)
of the full 10.3 $\mu$m flux of the star.  The resolved emission is detected
at a position angle of 164$\degr$, and assuming a Gaussian flux distribution
is responsible or the emission, has a diameter of about 20 AU.  Follow up
observations are needed to probe the geometry of the resolved emission.

\subsection{AB Aur}
\begin{figure}
\epsscale{0.85}
\plotone{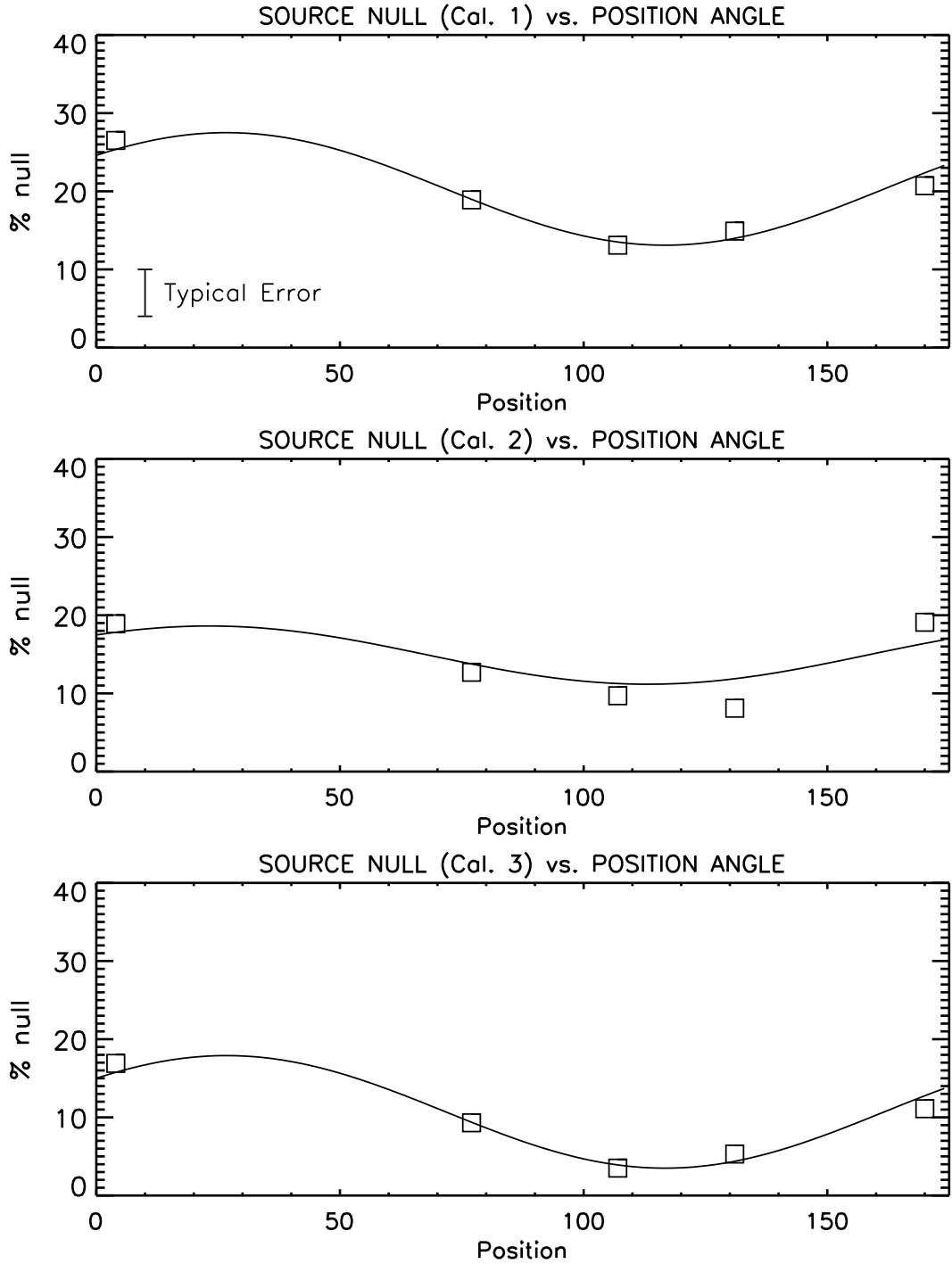}
\caption{Source nulls vs. rotation of interferometer baseline derived for
AB Aur for each of three calibrations described above.  The variation in
null is consistent with the presence of an inclined disk.}
\label{fig-abaur_source}
\end{figure}

Observations of AB Aur in 2004 were taken at five different rotations of
the interferometer baseline spanning $115\degr$ in increments of $30\degr$.
This rotation allows us to probe a range of position angles around
the star for resolved emission.  The source null for each rotation and
corresponding position angle is listed in Table \ref{tab-abaur_source}
for each of three different calibrations of the data (see next paragraph).
Plots of the data are shown in Figure \ref{fig-abaur_source}.  If
the emitting dust is in an inclined disk, one would expect the dependence
of the source null vs. position angle to be sinusoidal
\footnote{The transmitted signature of the nulling
interferometer is an interference pattern with interference fringes along the
baseline.  If these fringes are parallel to the major axis of a disk, more
of the disk's light will be nulled, resulting in a lower percentage of
remaining light.  When the fringes are aligned orthogonally to the major axis,
the value of the source null will be higher (i.e., there is more light
remaining).  Therefore, there should be a sinusoidal variation in null with
respect to rotation of the baseline.}.
From a fit to the data, of the form $N = a + b * sin(PA + \theta)$,
with the period fixed at 180$\degr$ \citep{liu03}, we derive physical properties
for an inclined disk around AB Aur.  For the fit, the values of $a$, $b$, and
$\theta$ are physically related to the size, inclination, and PA of the
major-axis, respectively.

Observations of the calibrator star, $\beta$ Gem, were taken before and
after the observations of AB Aur.  The calibrator shows
a significant change in the level of null we were able to achieve during
the observations, due probably to changes in observing conditions and
the effectiveness of AO wavefront correction.  As a result, we calibrate
our data (calculate the source nulls) for AB Aur in three different
ways to determine how this affects the results.  The different calibrations
are as follows: 1) We use the first calibrator measurement to calculate
all the source nulls; 2) We use a linear fit (in time) between the two
calibrator measurements to calculate the source nulls; 3) We use the last
calibration to calculate all the source nulls.

All three methods of calibration yield a result in which the dust distribution
is significantly more resolved in one PA ($\approx 30\degr$) than another
offset by $90\degr$ ($\approx 120\degr$).  This result is suggestive of a
flattened or elongated structure as the source of MIR excess emission.
Assuming two simple brightness distributions (a Gaussian disk and a ring),
physical parameters are derived from the fits to the source null vs.
position angle (Table \ref{tab-sinefit}).  For each model we have
derived the size and inclination of the material needed to reproduce the
null vs. position angle profile we have observed.  This is repeated for each
of the three calibrations, allowing us to assess the error introduced
by the calibration issues described above.  All three calibrations yield
similar results and indicate that the 10.3 $\mu$m emission originates from a
separation 12 to 17 AU from the star.  The presumed disk has a significant
inclination, 45 to 65 degrees from face-on and the position angle (PA) of the
major axis of the disk is $30\degr \pm 15\degr$.  If the actual distribution
of warm dust is a combination of a flattened structure and a uniform symmetric
component (such as a disk plus envelope), then the disk component would
need to be more inclined to account for the amplitude in null variation.

Previous studies have also observed AB Aur at MIR
wavelengths.  \citet{cj03} used the Keck I telescope to observe the star
at 11.7 and 18.7 $\mu$m.  They find that it is marginally resolved
at the longer wavelength.  At 18.7 $\mu$m, they find an angular diameter of
about 1$\arcsec$ at the half-maximum flux level.  This suggests that the
18 $\mu$m emission is originating from a separation of about 70 AU,
several times greater than the 10 $\mu$m emission.  A study by \citet{marsh}
finds evidence for resolved emission at 11.7 and 17.9 $\mu$m and
derives diameters of 40 and 80 AU for the emission, respectively.
Taking into account the derived size scales from this and both previous MIR
studies (separations of 12-17 AU, 20 AU, and 40-70 AU for the 10.3, 11.7,
and 18 $\mu$m emission, respectively), we note that the wavelength vs.
separation profile agrees with the radial temperature profile expected for
a flared disk, T $\sim$ r$^{-1/2}$ \citep{cg97}, assuming that the emission
is primarily thermal in nature.  We also note that the NIR
study of \citep{m-g01} finds that the thermal 2$\mu$m disk size is about
0.7 AU, which is also roughly consistent with a continuous flared disk.

Studies at other wavelengths include NIR studies (e.g.
\citet{f04,eis03}), and millimeter observations \citep{natta01,ms97},
\citet{ms97} found that AB Aur is "marginally resolved" in molecular line
emission at 3 mm.  They find a major axis PA of $79\degr$ with an inclination
of $76\degr$ from face-on. The significant inclination of
the disk agrees with the derived inclination of this study.  However
more recent studies in the millimeter \citep[and references therein]{natta01}
cite a much smaller inclination ($< 30\degr$).  A recent NIR
study by \citet{f04} using AO coronagraphic observations finds
a scattered light disk in H-band with a PA of $58\degr$ and a significantly
smaller inclination of $30\degr$ from face-on.  \citet{eis03} used the
Palomar Testbed Interferometer to obtain K-band observations of inner
($<0.5$ AU) disk surrounding AB Aur and find the inclination to be small,
within $30\degr$ of face-on, in agreement with \citet{f04}.  The observations
of this study suggest a greater inclination for the MIR emission.  One also
notes that the major axis PA derived for the MIR emission in this study
differs significantly (50-70\degr) from previous studies both in the NIR and
millimeter.  This points to a difference in geometry for the dust between
the inner (a few AU) and outer (hundreds of AU) system.  The biggest
difference between this study and those at other wavelengths is in the
inclination of the disk.  Previous studies at several different wavelengths
all agree on a significantly smaller inclination than found by this study.
This, in combination with the discrepancies in the PA of the disk suggest
that the structure may be more complex than a disk alone, where emission
at different wavelengths are dominated by material with a different
distribution.

Comparing AB Aur to a sample of 14 Herbig Ae stars observed in the MIR
by \citet{leinert} may be helpful in placing AB Aur into context and gaining
insight into their circumstellar environments.  The \citet {leinert} study finds
a correlation between the size of resolved emission and the SED classifications of
\citet{meeus}.  They find that the Meeus et al. Type I sources, characterized
by a rising MIR SED, tend to have spatially larger circumstellar emission
regions.  By contrast, Type II sources have flat or declining MIR SEDs have
smaller resolved sizes.  The resolved size of the AB Aur emission found
in this study would classify AB Aur as a Type I source, consistent with the
initial classification by \citet{meeus} by SED alone.  While it appears that
the each type shows similar physical characteristics, there also
seems to be evidence that a simple, all encompassing physical model may not
be an ideal explanation for each Herbig type.  For example, it is interesting to
note some differences, from nulling observations and other previous
studies, between AB Aur and another resolved Type I Herbig Ae star, HD 100546.
The primary difference highlighted by nulling observations is that AB Aur's
radial wavelength (temperature) profile seems to be consistent with a continuous
disk, whereas HD 100546's relative 10 $\mu$m vs. 20 $\mu$m emission region sizes
suggest an inner clearing \citep{liu03}.  Other studies have found difference
in the age (10 Myr for HD 100546 vs. 2-5 Myr for AB Aur; \citet{vdA98,ms97}) and
evolutionary states \citep{bouwman} of the two stars.  It seems, therefore,
that although the two stars show similarities in the size of their resolved
emission at 10 $\mu$m, the emission arises from physically different
distributions of circumstellar dust.

\section{Ongoing Work: Surveys of Intermediate-Mass Stars}
The observations of AB Aur and V892 Tau presented here are part of a
survey of 14 nearby HAE stars for resolved circumstellar material in the
MIR.  Nulling interferometric observations
with the BLINC-MIRAC instrument from the MMT and Magellan I (Baade) 6.5-m
telescopes are now complete.  Results and an analysis of the full sample will
be presented in an upcoming paper.  Also currently underway is a survey of
nearby intermediate-mass main-sequence stars for second-generation exozodiacal
dust, again utilizing nulling interferometry with AO.  Completed
observations include those of Vega, presented in \citet{liu04}.

\section{Acknowledgements}
W.L. was supported under a Michelson Graduate Fellowship.  The authors thank
the staff at the MMT for excellent support.  We thank B. Duffy for technical
support with the BLINC-MIRAC instrument.  We thank the anonymous referee
for helpful comments.  BLINC was developed under a grant
from NASA/JPL, and MIRAC is supported by the NSF and SAO.  The MMT AO system
was developed with support from AFOSR.

\clearpage
\begin{deluxetable}{cccccc}
\tablecaption{Source Nulls for AB Aur \label{tab-abaur_source}}
\tablewidth{0pt}
\tablehead{
\colhead{Set \#} &
\colhead{Rotation} &
\colhead{Maj. Axis Pos.($\degr$ E of N)} &
\colhead{Cal. 1} &
\colhead{Cal. 2} &
\colhead{Cal. 3}
}
\startdata
1 & -160 & 4 & 26.5\%  & 18.9\% & 16.9\%\\
2 & -135 & 170 & 20.7 & 19.1 & 11.1\\
3 & -105 & 131 &14.9 & 8.1 & 5.3\\
4 & -75 & 107 & 13.1 & 9.7 & 3.5\\
5 & -45 & 77 & 18.9 & 12.7 & 9.3\\
\enddata
\end{deluxetable}

\clearpage
\begin{deluxetable}{ccccc}
\tablecaption{Derived Parameters for AB Aur Disk \label{tab-sinefit}}
\tablewidth{0pt}
\tablehead{
\colhead{Cal. Set} &
\colhead{Guassian FWHM (AU)} &
\colhead{Incl.($\degr$)} &
\colhead{Ring Diameter (AU)} &
\colhead{Incl.($\degr$)}
}
\startdata
1 & $30 \pm 3$ & $47 \pm 5$ & $34 \pm 3$ & $45\pm 5$\\
2 & $27 \pm 3$ & $52 \pm 5$ & $30 \pm 3$ & $50\pm 5$\\
3 & $24 \pm 2$ & $64 \pm 6$ & $28 \pm 3$ & $63\pm 6$\\
\enddata
\end{deluxetable}

\clearpage


\begin{thebibliography}

\bibitem[Bouwman et al.(2003)]{bouwman}
Bouwman, J., de Koter, A., Dominik, C., \& Waters, L.~B.~F.~M.\ 2003, \aap,
401, 577
\bibitem[Chen \& Jura(2003)]{cj03} Chen, C.~H.~\& Jura, M.\
2003, \apj, 591, 267
\bibitem[Chiang \& Goldreich(1997)]{cg97} Chiang, E.~I.~\&
Goldreich, P.\ 1997, \apj, 490, 368
\bibitem[Hartmann, Kenyon, \& Calvet(1993)]{hart93} Hartmann,
L., Kenyon, S.~J., \& Calvet, N.\ 1993, \apj, 407, 219
\bibitem[Eisner et al.(2003)]{eis03} Eisner, J.~A., Lane,
B.~F., Akeson, R.~L., Hillenbrand, L.~A., \& Sargent, A.~I.\ 2003, \apj,
588, 360
\bibitem[Elias(1978)]{elias} Elias, J.~H.\ 1978, \apj, 224,
857
\bibitem[Fukagawa et al.(2004)]{f04} Fukagawa, M., et al.\
2004, \apjl, 605, L53
\bibitem[Grady et al.(1999)]{grady} Grady, C.~A., Woodgate,
B., Bruhweiler, F.~C., Boggess, A., Plait, P., Lindler, D.~J., Clampin, M.,
\& Kalas, P.\ 1999, \apjl, 523, L151
\bibitem[Haas, Leinert, \& Richichi(1997)]{haas97} Haas, M.,
Leinert, C., \& Richichi, A.\ 1997, \aap, 326, 1076
\bibitem[Hillenbrand et al.(1992)]{hill92}
Hillenbrand, L.~A., Strom, S.~E., Vrba, F.~J., \& Keene, J.\ 1992, \apj,
397, 613
\bibitem[Hinz(2001)]{hinz_phd} Hinz, P.~M.\ 2001, Ph.D.~Thesis, University of
Arizona
\bibitem[Hoffmann et al.(1998)]{hoff} Hoffmann, W.~F., Hora,
J.~L., Fazio, G.~G., Deutsch, L.~K., \& Dayal, A.\ 1998, \procspie, 3354,
647
\bibitem[Kenyon \& Hartmann(1987)]{kh87} Kenyon, S.~J.~\&
Hartmann, L.\ 1987, \apj, 323, 714
\bibitem[Lada \& Adams(1992)]{la92} Lada, C.~J.~\& Adams,
F.~C.\ 1992, \apj, 393, 278
\bibitem[Leinert et al.(2004)]{leinert} Leinert, C., et al.\
2004, \aap, 423, 537
\bibitem[Liu et al.(2004)]{liu04} Liu, W.~M., et al.\ 2004,
\apjl, 610, L125
\bibitem[Liu et al.(2003)]{liu03} Liu, W.~M., Hinz, P.~M.,
Meyer, M.~R., Mamajek, E.~E., Hoffmann, W.~F., \& Hora, J.~L.\ 2003, \apjl,
598, L111
\bibitem[Mannings \& Sargent(1997)]{ms97} Mannings, V.~\&
Sargent, A.~I.\ 1997, \apj, 490, 792
\bibitem[Marsh et al.(1995)]{marsh} Marsh, K.~A., Van Cleve,
J.~E., Mahoney, M.~J., Hayward, T.~L., \& Houck, J.~R.\ 1995, \apj, 451,
777
\bibitem[Meeus et al.(2001)]{meeus} Meeus, G., Waters,
L.~B.~F.~M., Bouwman, J., van den Ancker, M.~E., Waelkens, C., \& Malfait,
K.\ 2001, \aap, 365, 476
\bibitem[Millan-Gabet et al.(2001)]{m-g01}
Millan-Gabet, R., Schloerb, F.~P., \& Traub, W.~A.\ 2001, \apj, 546, 358
\bibitem[Miroshnichenko et al.(1999)]{miro99} Miroshnichenko, A.,
Ivezi{\' c} , {\v Z}., Vinkovi{\' c} , D., \& Elitzur, M.\ 1999,
\apjl, 520, L115
\bibitem[Natta et al.(2001)]{natta01} Natta, A., Prusti, T.,
Neri, R., Wooden, D., Grinin, V.~P., \& Mannings, V.\ 2001, \aap, 371, 186
\bibitem[van den Ancker et al.(1998)]{vdA98} van den Ancker, M.~E.,
de Winter, D., \& Tjin A Djie, H.~R.~E.\ 1998, \aap, 330, 145

\end{thebibliography}
\end{document}